\documentclass[twocolumn, preprintnumbers, 
endnote,nofootinbib,prl,9pt,superscriptaddress]{revtex4}

\usepackage{epsfig,subfigure}

\usepackage{epstopdf}

\usepackage[utf8]{inputenc}
\usepackage{graphicx}
\usepackage{amssymb}
\usepackage{amsxtra}
\usepackage{amsmath}
\usepackage{booktabs,multirow,tabularx}
\usepackage{slashed}
\usepackage{float}
\usepackage{placeins}
\usepackage{rotating}
\usepackage{lscape}
\usepackage{color}
\usepackage{hyperref}

\usepackage[Q=yes,pverb-linebreak=no]{examplep}

\DeclareGraphicsExtensions{.eps}
\graphicspath{{./}}

\newcommand{\vev}[1]{\langle {#1} \rangle}
\newcommand{\lsim}{\lesssim}
\newcommand{\gsim}{\gtrsim}

\newcommand{\gev}{\,\textrm{GeV}}

\newcommand{\tev}{\,\textrm{TeV}}

\newcommand{\ord}[1]{\mathcal{O}{(#1)}}

\def\beq{\begin{equation}}
\def\bea{\begin{eqnarray}}
\def\eeq{\end{equation}}
\def\eea{\end{eqnarray}}
\def\beqnl{\begin{align}}
\def\endal{\end{align}}

\newcommand{\sute}{SU(2)_e}
\newcommand{\DBL}{\Delta_{B-L}}

\DeclareFontFamily{U}{cbgreek}{}
\DeclareFontShape{U}{cbgreek}{m}{n}{
        <-6>    grmn0500
        <6-7>   grmn0600
        <7-8>   grmn0700
        <8-9>   grmn0800
        <9-10>  grmn0900
        <10-12> grmn1000
        <12-17> grmn1200
        <17->   grmn1728
      }{}
\DeclareFontShape{U}{cbgreek}{bx}{n}{
        <-6>    grxn0500
        <6-7>   grxn0600
        <7-8>   grxn0700
        <8-9>   grxn0800
        <9-10>  grxn0900
        <10-12> grxn1000
        <12-17> grxn1200
        <17->   grxn1728
      }{}

\DeclareRobustCommand{\qoppa}{%
  \text{\usefont{U}{cbgreek}{\normalorbold}{n}\symbol{19}}%
}
\DeclareRobustCommand{\Qoppa}{%
  \text{\usefont{U}{cbgreek}{\normalorbold}{n}\symbol{21}}%
}
\makeatletter
\newcommand{\normalorbold}{%
  \ifnum\pdf@strcmp{\math@version}{bold}=\z@ bx\else m\fi
}
\makeatother

\begin{document}

\title{\boldmath Asymmetric Dark Matter in Extended Exo-Higgs Scenarios}

\author{Hooman Davoudiasl\footnote{email: hooman@bnl.gov}
}

\author{Pier Paolo Giardino\footnote{email: pgiardino@bnl.gov}
}
\affiliation{Department of Physics, Brookhaven National Laboratory,
Upton, NY 11973, USA}

\author{Cen Zhang\footnote{email: cenzhang@ihep.ac.cn}
}

\affiliation{Department of Physics, Brookhaven National Laboratory,
Upton, NY 11973, USA}
\affiliation{Institute of High Energy Physics, Chinese Academy of Sciences,
Beijing, 100049, China}

\begin{abstract}

The exo-Higgs model can accommodate a successful baryogenesis mechanism that closely mirrors electroweak baryogenesis in the Standard Model, but avoids its shortcomings.  We extend the exo-Higgs model by the addition of a singlet complex scalar $\chi$.  In our model, $\chi$ can be a viable asymmetric dark matter (ADM) candidate.  
We predict the mass of the ADM particle to be $m_\chi\approx1.3 \gev$.
The leptophilic couplings of $\chi$ can provide for efficient annihilation of 
the ADM pairs.
We also discuss the LHC signals of our scenario, and in particular the production and decays of exo-leptons which would lead to ``lepton pair plus missing energy" final states. Our model typically predicts potentially detectable gravitational waves originating from the assumed strong first order phase transition at a temperature of $\sim$ TeV. If the model is further extended to include new heavy vector-like fermions, {\it e.g.} from an ultraviolet extension, $\chi$ couplings could explain the $\sim 3.5\sigma $ muon $g-2$ anomaly.   

\end{abstract}

\maketitle

\section{Introduction\label{sec:intro}}

In  Ref.~\cite{exo-Higgs}, we introduced the {\it exo-Higgs} $\eta$ associated with breaking a new $SU(2)_e$ gauge interaction that we dubbed {\it exo-spin} [since the Standard Model (SM) particles do not carry $SU(2)_e$ charge and this force is {\it outside} the SM].  The model included fermions, exo-quarks $\Qoppa$ and exo-leptons $\Lambda$, that can be assigned SM  baryon ($B$) and lepton ($L$) numbers, respectively.   The quantum numbers of the exo-fermions were chosen in a way that leads to a $B-L$ anomaly under $SU(2)_e$ gauge interactions.  Therefore, by choosing the parameters of the model to ensure a first order phase transition, the model could potentially accommodate the generation of a $B-L$ asymmetry $\DBL$ in the early universe, through the action of $\sute$ exo-sphalerons.  This $B-L$ asymmetry would not be washed out by the SM sphalerons and can hence remain to become the baryon asymmetry of the universe (BAU).   

The above {\it exo-baryogenesis} scenario could be realized in close analogy with how baryogenesis {\it could have} worked in the SM, if the electroweak phase transition were strongly first order and $CP$ violating effects were more than $\sim 10^{10}$ larger (see Ref.~\cite{Gavela:1994dt}, for example).  However, the exo-spin sector in Ref.~\cite{exo-Higgs} can potentially accommodate the requisite first order phase transition and significant $CP$ violating effects to yield successful generation of the observed BAU.  

In this work, we minimally extend the exo-Higgs model \cite{exo-Higgs} by adding a singlet complex scalar $\chi$ which carries a good global symmetry.  
The addition of $\chi$ removes {\it ad hoc} mass scales that appeared in the original exo-Higgs model and replaces them with 
the Yukawa couplings of $\chi$ to charged exo-leptons and SM leptons.  This scalar can be stable and provide a viable 
asymmetric dark matter (ADM) candidate \cite{Kaplan:2009ag,Davoudiasl:2012uw,Petraki:2013wwa,Zurek:2013wia}.  We will show that, at typical strengths, the new Yukawa couplings of $\chi$ can mediate the annihilation of $\chi \chi^*$ symmetric population in the early Universe, which is a requirement for establishing the correct ADM abundance.  The above new Yukawa coupling leads to prompt decays of the exo-leptons into SM leptons.  These decays transfer the high scale $B-L$ asymmetry generated during the $SU(2)_e$ phase transition to the SM sector (see also Ref.~\cite{Petraki:2011mv} for a proposal which includes some ingredients similar to those in our work).

Since the Yukawa couplings of $\chi$ to exo-leptons and charged leptons are responsible for both the transfer of $\Delta_{B-L}$ {\it and} effective annihilation of $\chi\chi^*$ pairs, as mentioned before, at least one of the exo-leptons $\Lambda$ is required to have 
a mass $\lsim 1$~TeV.  Given the requisite non-zero electric 
charge of the exo-leptons, their production at the LHC can go through Drell-Yan processes 
and lead to striking large-missing-energy-plus-charged-lepton signals.  Hence, key 
interactions involved in our baryogenesis proposal, namely the aforementioned Yukawa couplings 
of $\chi$,  may potentially be accessible at the LHC, with a few 100~fb$^{-1}$ of data.  We also note that, as already pointed out in our previous paper \cite{exo-Higgs}, a typical prediction of our model is potentially detectable gravitational waves at future space-based observatories (see for example Ref. \cite{Grojean:2006bp,Schwaller:2015tja}). These features of our work provide direct probes of important hypothesized early Universe processes, very much in the spirit of the original electroweak 
baryogenesis proposals.  We also add that our model yields a definite prediction for the mass 
of the DM state $\chi$, however this feature is shared with various other ADM proposals.

An interesting question is whether the $\sim 3.5 \sigma$ 
deviation of the measured muon anomalous magnetic moment, $g_\mu-2$, from the SM prediction \cite{PDG} may be due to the above couplings of the ADM candidate $\chi$ to charged leptons.  We find that the above setup will not provide the requisite $g_\mu-2$ contributions without large couplings $\gg 1$, which would not allow reliable estimates. However it is still possible to explain the deviation in our framework, once we assume the existence of a new scale of physics above $\sim$ few TeV. While this assumption could seem gratuitous, it is actually a well-motivated  extension of our base model if, as detailed below, we assume that scalar mass parameters originate from higher scale symmetry breaking.  In that case, one could naturally assume that the exo-Higgs potential scale of a few TeV is set from a nearby symmetry breaking around $\ord{10}$~TeV.  One may then expect additional fields, in particular new heavy fermions, whose mass is endowed by the higher scale symmetry breaking.  With a minimal addition of such fermions at $\sim$ few TeV, our model can also account for the $g_\mu-2$ anomaly.  We will provide a simple extension to that end in the appendix.  Hence, in the context of ultra violet extended exo-Higgs models, the $g_\mu-2$ deviation may be the first non-gravitational signal of dark matter.

Next we will briefly outline the main ingredients of the exo-Higgs model and minimal extensions of it mentioned above.

\section{A Minimal Extension of the Exo-Higgs Model}

The exo-Higgs model was introduced in detail in Ref.~\cite{exo-Higgs}.  Here, we only review its main ingredients and features that will be relevant to this work.  The model introduces 
a new $SU(2)_e$ gauge symmetry, exo-spin, under which all SM particles are singlets, whose gauge bosons we denote with the symbol $\omega$.  The new matter content includes a scalar doublet $\eta$, whose vacuum expectation value (vev) $\vev{\eta}\equiv v_\eta/\sqrt{2}$ breaks $\sute$, and three generations of exo-quarks $\Qoppa$ and exo-leptons $\Lambda$, whose $SU(2)_e\otimes SU(3)_c \otimes SU(2)_L \otimes U(1)_Y$ quantum numbers are given by 
\bea
\Qoppa_L=(2,3,1,-\frac{1}{3})\;\; &;& \;\; 2\times \Qoppa_R=(1,3,1,-\frac{1}{3}) \\ \nonumber
2\times \Lambda_L=(1,1,1,-1)\;\; &;& \;\;  \Lambda_R=(2,1,1,-1)\,,
\label{exo-fermions}
\eea
in the implied order. We assign baryon $B=\frac{1}{3}$ and lepton $L=1$ numbers to exo-quarks and exo-lepton, respectively.  The above 
charges then lead to a $B-L$ anomaly under $\sute$. 

By complete analogy with the SM, one can write down various kinetic terms, a potential for $\eta$, and Yukawa couplings for 
$\Qoppa$ and $\Lambda$ fields whose mass will be proportional to $v_\eta$.  The exo-sector and the SM can be coupled through the Higgs 
portal and Higgs-$\eta$ mixing, as well as mixed Yukawa terms, for $\Qoppa$ and SM quarks $q$, of the type $H \bar{q}_L \Qoppa_R$ and 
$\eta \,\bar{\Qoppa}_L q_R $.  The symmetries of the model do not allow mixed Yukawa terms for $\Lambda$ and SM leptons $\ell = e,\mu,\tau$.  However, 
one can write down mass terms of the type $m \bar{\Lambda}_L \ell_R$, with the mass scale $m$ subject to phenomenological constraints, but otherwise 
{\it ad hoc}.   

We will extend the above setup by introducing a new field, a complex scalar $\chi$ that carries a good global charge $Q_\chi = +1$.  We also 
demand that $\Lambda_{L,R}$ both have $Q_\chi = +1$. This assignment of charges forbids the above $\Lambda$-$\ell$ mass mixing term, 
but one can now write down a new Yukawa coupling of the type 
\beq
\lambda_\ell\, \chi \bar{\Lambda}_L \ell_R.  
\label{chiYukawa}
\eeq
We will assume that $\chi$ is the lightest 
$Q_\chi\neq 0$ state, and so it will be a stable particle and a potential DM candidate.  For the sake of completeness, we also mention 
that as usual, one can add the new quartic interactions 
\beq
\lambda_\chi (\chi^\dagger\chi)^2 + 2 k_{\chi H} \chi^\dagger \chi H^\dagger H + 
2 k_{\chi \eta} \chi^\dagger \chi \eta^\dagger \eta \,,
\label{quartic} 
\eeq
to the scalar potential.  The mixed terms can in principle supply the required mass term for $\chi$, 
after exo-spin and electroweak symmetry breaking.   

We recall that in the exo-Higgs scenario, baryogenesis 
proceeds through the generation of a $B-L$ asymmetry during a first order $\sute$ phase transition \cite{exo-Higgs}.  That asymmetry initially resides 
in the exo-fermions, but is then transferred to the SM quarks and leptons in prompt decays.  In particular, $\Lambda \to \chi \, \ell$ 
will release lepton number into the $\sute$ broken phase, which will eventually result in $B\neq 0$ through the action of the SM sphalerons.  However, 
this decay will also inject the net $Q_\chi$ charge of $\Lambda$ into the electroweak plasma through the $\chi$ final state. We would 
then end up with a net asymmetry in $\chi$ particles, as $Q_\chi$ conservation forbids their decays.  

In the above scenario, if $\chi \chi^*$ pairs 
annihilate efficiently, the remaining $\chi$ asymmetry can supply the observed energy density of DM.  Since the number density 
of $\chi$ ADM and baryons are tied in our model, for $\chi$ to have the correct contribution to the cosmic energy budget it is required to have 
a particular mass $m_\chi$.  To predict the required value of $m_\chi$, we would then need to know the relation between baryon and $\chi$ asymmetries, 
which we will derive next.

\section{The Relation Between \boldmath $\Delta Q_\chi$ and $\Delta B$}

We will assume that all the interactions in the exo-sector are in thermal equilibrium above the 
temperature $T_c^e \sim v_\eta$ of $\sute$ phase transition.  This is a mild assumption that can be 
quite typically realized in our model.  Given the similarity of the exo-Higgs model and the SM, the calculation of 
the relation between various fermion numbers and charges closely follows that of the SM sector, as detailed in Ref.~\cite{H&T}.  
In the unbroken phase, the chemical potentials of the gauge bosons must be zero. Also Eq. (4) of Ref.~\cite{exo-Higgs} yields a zero chemical potential for $\eta$ (the chemical potentials of $\eta$ and $\tilde{\eta}$ are equal and opposite).
Thus the chemical potentials of $\qoppa_L$ and $\qoppa_R$ and of $\qoppa$'s in a exo-doublet are the same and will be denoted by the symbol $\mu_\qoppa$. A similar result applies to $\Lambda$'s and their chemical potential $\mu_\Lambda$.
This leaves us with only the following relations:
\bea
&&\mu_{dR}=\mu_\qoppa, \ \ \ \ \mu_{uL}=\mu_\qoppa+\mu_0, \\
&&\mu_{iR}=\mu_\Lambda - \mu_\chi,\ \ \ \ \ 3 \mu_\qoppa-\mu_\Lambda=0, \nonumber \\
&&\sum_i(\mu_{iR}+\mu_{iL}) + 3(\mu_{dR}+\mu_{dL})-6(\mu_{uR}+\mu_{uL})\nonumber \\ &&+12(\mu_\Lambda+\mu_\qoppa)-2\mu_0=0, \nonumber
\label{eq:chem}
\eea
where we follow the notation of Ref.~\cite{H&T} for the SM particles chemical potentials, $\mu_\chi$ is the chemical potential of $\chi$ and the last two equations are respectively the exo-sphaleron and the conservation of the electric charge. After solving the system obtained from Eq. (\ref{eq:chem}) and Eqs. (2) and (3) of Ref.~\cite{H&T}, it is easy to obtain $\Delta_{B-L}$ and $\Delta Q_\chi$ in terms of the $\qoppa$ chemical potential:
\beq
\Delta_{B-L}=\frac{789}{19} \mu_\qoppa,\ \ \ \ \Delta Q_\chi=\frac{1008}{19} \mu_\qoppa.
\eeq
We treat the electroweak transition as a smooth cross over which implies that the SM sphalerons are in thermal equilibrium after 
the Higgs gets a vev $v_H \equiv \sqrt{2}\vev{H}$ in the early Universe. Thus the relation between $\Delta_{B-L}$ and $\Delta B$ is the one in Eq. (11) of Ref.~\cite{H&T}, and we find 
\beq
\frac{\Delta Q_\chi}{\Delta B} = \frac{1036}{263}.
\label{QchiB}
\eeq

The ratio of DM and baryonic energy densities is \cite{PDG}
\beq
\frac{\Omega_{\rm DM}}{\Omega_B} \approx 5.3  
\label{OmegaDM}
\eeq
which, assuming that the $\chi$ asymmetry accounts for all DM, 
yields 
\beq
m_\chi \approx 1.3~{\rm GeV}.
\label{mchi}
\eeq
The above value of DM mass $m_\chi$ is a direct prediction of our scenario, with important implications for the viability of 
the assumed setup. 

\section{Conjugate Pair Annihilation of DM}

To link the ADM number density with that of baryons, we need to make sure that the 
symmetric DM-anti-DM population is efficiently depleted.  In our model, so far, there are only two ways for 
$\chi \chi^*$ pairs to annihilate: through the Yukawa coupling 
(\ref{chiYukawa}) or else via interactions with Higgs or $\eta$ in Eq.~(\ref{quartic}).  
Since we have $m_\chi \ll v_\eta\,, v_H$, the quartic $\chi^\dagger \chi H^\dagger H$ and 
$\chi^\dagger\chi \eta^\dagger \eta$ interaction must be suppressed, otherwise $m_\chi^2$  would receive large contributions 
without tuning, once $H$ and $\eta$ condense.

Hence, the Yukawa couplings of $\chi$ are the only candidates for efficient annihilations into SM leptons, through $t$-channel Feynman diagrams.  Since the $\tau$ lepton is heavier than $\chi$, $e$ and $\mu$ are the only feasible 
final states. We have checked, using the {\sc MadDM} package \cite{Backovic:2013dpa}, that the interaction in Eq.~(\ref{chiYukawa}), with 
$\lambda_\ell \sim 1$ and $m_\Lambda \sim 1-2$~TeV can provide a sufficiently large 
cross section for efficient annihilation of $\chi$ conjugate pairs.   
 
The above possibility is not ruled out by the measured values of the muon and electron $g-2$ \cite{PDG}, as explained next. In fact, due to the interaction Eq.~(\ref{chiYukawa}), the $a_\ell\equiv (g_\ell-2)/2$ of a lepton $\ell$ receives a contribution from the exo-leptons $\Lambda$ given by \cite{Leveille:1977rc} 
\beq
\delta a_\ell=2\frac{m_\ell^2 \lambda_\ell^2}{16\pi^2}\int^1_0dx \frac{x^2-x^3}{\mathcal{D}},
\label{g-2}
\eeq 
where $\mathcal{D}\equiv m_\ell^2x^2+(m_\Lambda^2-m_\ell^2)x+m_\chi^2(1-x)$ and the 2 is due to the contribution of $\Lambda$'s in the same exo-doublet.
In the limit $m_\Lambda \gg m_\chi \gg m_\ell$ Eq. (\ref{g-2}) becomes 
\beq
\delta a_\ell\simeq\frac{\lambda_\ell^2 m_\ell^2}{48 \pi^2 m_\Lambda^2}.
\label{g-2lim}
\eeq 
In Eqs.~(\ref{g-2}) and (\ref{g-2lim}) we have assumed that the interaction in Eq.~(\ref{chiYukawa}) is flavor diagonal and that the exo-leptons in the same $SU(2)_e$ doublet have degenerate masses. However, these assumptions can be easily lifted provided that the flavor off-diagonal terms are small enough to avoid bounds on lepton flavor changing processes like $\mu\to e\gamma$.  We see that for $\lambda_\ell\sim 1$ and $m_\Lambda\sim 1$~TeV 
the contributions to $g_\ell-2$ are too small to provide constraints.

As mentioned in the introduction, we can accommodate the present difference between the measured and calculated value of $a_\mu$, that is $\delta a_\mu = 276(80) \times 10^{-11}$ \cite{PDG,Kurz:2014wya,Chen:2015vqy}, by introducing additional heavy degrees of freedom at or above a few TeV.  
The reason the Yukawa coupling in Eq.~(\ref{chiYukawa}) yielded a small contribution is due to a cancellation between scalar and pseudo-scalar contributions to $\delta a_\mu$ from that 
operator \cite{Leveille:1977rc}.  To change this result, one must induce a new coupling 
\beq
\bar\lambda_\ell \chi \, \bar{\Lambda}_R \,\ell_L
\label{lambar}
\eeq
where we have implicitly assumed that the relevant fields represent mass eigenstates after electroweak symmetry breaking.  One can then show that in this case the contribution to $g-2$ is 
\bea
\delta a_\ell&=&\frac{m_\ell^2 (\lambda_\ell^2+\bar\lambda_\ell^2)}{8\pi^2}\int^1_0dx \frac{x^2-x^3}{\mathcal{D}}\nonumber \\
&+&\frac{m_\ell m_\Lambda \lambda_\ell\bar\lambda_\ell}{4\pi^2}\int^1_0dx \frac{x^2}{\mathcal{D}}.
\label{g-2:op}
\eea 

We see that the second term in Eq.~(\ref{g-2:op}) provides the dominant contribution, and 
in the limit $m_\Lambda \gg m_\chi \gg m_\ell$ Eq.(\ref{g-2:op}) yields
\beq
\delta a_\ell\simeq\frac{\lambda_\ell\bar\lambda_\ell m_\ell}{8 \pi^2 m_\Lambda}.
\label{g-2lim:op}
\eeq 
The above contribution can have the correct size to address the muon $g-2$ anomaly if $\bar \lambda_\ell \sim 10^{-3}$, for $v_\eta$ and $m_\Lambda$ at $\mathcal{O}$(TeV).

Schematically, the interaction (\ref{lambar}) is induced by physics at a scale $\mathcal{M}$, generating the dimension-6 operator 
\beq
\frac{\chi (\eta \bar{\Lambda}_R) (H^\dagger l_L) }{\mathcal{M}^2},
\label{op}
\eeq
where $l_L$ is a SM lepton doublet. Hence, 
$\bar\lambda_\ell=v_\eta v_H/(2\mathcal{M}^2)$ and requiring $\bar\lambda_\ell \sim 10^{-3}$ yields 
$\mathcal{M}\sim \mathcal{O}$(10\,TeV), which suggests that new states at a few TeV 
with reasonable couplings can yield the requisite contribution to $a_\mu$.  We will elaborate further on this point in the appendix, where we will give an explicit example of a UV completion.

\section{LHC phenomenology}

In this section, we are interested in exploring the experimental signatures of our model.  
To this end it is convenient to fix a benchmark scenario for the exo-sector:
\begin{flalign}
\begin{array}{lcll}
m_\eta & = & 1.5 \tev & \text{Mass of the $\eta$ field}\\
v_\eta & = & 2.5 \tev & \text{Vev of the $\eta$ field}\\
m_\Qoppa^h & = & 1.5 \tev & \text{Mass of the heaviest $\Qoppa$}\\
m_\Qoppa^l & \sim & 1 \tev & \text{Mass of the lightest $\Qoppa$'s}\\
m_\Lambda & = & 1 \tev & \text{Mass of $\Lambda$'s}\\
 g_e & = & 2 & \text{$SU(2)_e$ gauge coupling},\\
\end{array}
\label{eq:benchmark}
\end{flalign}
where we have assumed a mass hierarchy between one $\Qoppa$ flavor and the other two. 
We set the parameters in Eq.~(\ref{eq:benchmark}) to ensure that the phase transition is strongly first order, namely $3 g_e^3/(16 \pi\lambda_\eta)\gsim1$ where $\lambda_\eta$ is the coefficient of the quartic term $(\eta^\dagger \eta)^2$; see Ref.~\cite{exo-Higgs} for a discussion of 
a first order phase transition in the exo-Higgs scenario and Ref.~\cite{Quiros:1999jp} for details.  
With these parameters, the model maintains stability and perturbativity up to very high scales ($\sim 10^{5-6} \tev$).
In particular, the rather large value of $g_e$ assumed in the above 
benchmark scenario does not lead to a loss of perturbativity near the 
scales of interest in our work.  To see this, we note that $\alpha_e 
\equiv g_e^2/(4 \pi) = 1/\pi$, which is sufficiently small to allow a 
perturbative analysis.  Also, one can easily check that the running of 
$g_e$ is governed by a beta-function that is numerically equal to that of 
the SM $SU(2)_L$, which is asymptotically free.  Hence, $SU(2)_e$ 
interactions are perturbative at the TeV scale and become weaker at larger 
scales.

The SM mixes with the exo-sector through a Lagrangian term of the form
\bea
\label{eq:Lm}
\mathcal{L}_m&=&  2 k_{\eta H} \eta^\dagger\eta H^\dagger H - Y_{\Qoppa q} \eta\, \bar{\Qoppa}_L d_R - Y_{q\Qoppa} H \bar{q}_L  \Qoppa_R,
\eea
that could modify the decay rates of the Higgs into gluons, photons and $Z$'s 
through loops of $\Qoppa$'s and $\Lambda$'s, the couplings of the Higgs, $Z$ and $W$ to quarks, and induce FCNC operators in the quark sector.
However, we require that the breaking of the EW symmetry is a direct 
consequence of $\eta$ acquiring a vev, {\it i.e.} $k_{\eta H} v_\eta^2 = \mu_H^2 = m_H^2/2$, thus the mixing angle between $\eta$ and $H$ is
\beq
\tan(2\theta_{\eta H}) = \frac{4 k_{\eta H}\, v_H\, v_\eta}{m_\eta^2-m_H^2},
\label{thetaetaH}
\eeq
which yields $\theta_{\eta H}\sim 7\times 10^{-4}$.

The values of $Y_{\Qoppa q}$ and $Y_{q\Qoppa}$ in Eq.~(\ref{eq:Lm}) can also be easily set to be small, and the strongest constraint comes from the assumption that the interactions $\Qoppa\leftrightarrow q \eta$ and $\Qoppa\leftrightarrow q H$ are in thermal equilibrium at $T_c^e$.
In particular we find that the corresponding mixing angle between $\Qoppa$'s and quarks, $\theta_{\Qoppa, q}$, can be set as small as $\theta_{\Qoppa, q}\sim 10^{-4}$.
The $\Qoppa$'s  play a fundamental role in the baryogenesis since they provide a viable source of $CP$ violation, if there is a sufficiently large hierarchy between the masses of different generations of $\Qoppa$'s, and if the mixing matrix has large enough mixing angles. In particular, depending on the size of the mixing angles in the exo-CKM, a difference of $\mathcal{O}(50 \gev)$ between the masses of the lightest $\Qoppa$'s would be sufficient.  As shown in Ref.~\cite{exo-Higgs}, $\Qoppa$'s can be looked for in standard vector-like quark searches at LHC (see for example Ref.~\cite{Khachatryan:2015gza}).
The current limit on charge -1/3 vector-like quark is around $700\sim800$ GeV
\cite{Aad:2015gdg,Aad:2014efa,Aad:2015mba,Aad:2015kqa,Aad:2015tba,ATLAS:2016sno}.

Unlike in Ref.~\cite{exo-Higgs}, the exo-leptons $\Lambda$ are mostly produced
in pairs through Drell-Yan and can decay only through the process
$\Lambda\to\chi\ell$.  For $\Lambda\sim 1\tev$, the signal would be a pair of
opposite-sign-same-flavor\footnote{We remind the reader that in this setting we
	are imposing that the interaction in Eq.~(\ref{chiYukawa}) is flavor
diagonal.} leptons, with $p_T\gtrsim$ a few hundred GeV, and a large missing
$E_T$ (MET), which is hard to be missed.  Background processes in the SM mainly come
from the decay of the top quark and the weak gauge bosons, where the neutrinos
in the decay products may carry a large MET.  For this reason we
expect the dominant
backgrounds to come from $t\bar t$ production, $W$ pair production, and $tW$
associated production channels, while those from $ZZ$ and $Z\gamma^*$ can be
removed by requiring that the dilepton mass differ from the $Z$ boson, as
demonstrated in an analysis with the same final state
\cite{Khachatryan:2014qwa}.  To briefly estimate the potential sensitivity, we
define the signal region by a set of kinematic cuts defined in
Table~\ref{tab:cuts}.  The cross sections after each cut are also displayed.
From the results we find that at the $13 \tev$ LHC with $100$ fb$^{-1}$ we can
bring the background down to about one event, while still having
$\mathcal{O}(10)$ signal events.  While a more sophisticated study, possibly
including detector level simulation is required to determine the discovery
potential at the LHC precisely, we do not expect that our conclusion will be
changed significantly.

\begin{table*}
	\begin{tabular}{lccccc}
		\hline
		&Total & $p_{T}(l_1)>300$ GeV & $p_{T}(l_2)>200$ GeV &
		MET$>300$ GeV & $p_{T}(j)<80$ GeV
		\\
		\hline
		Signal & 0.18 & 0.18 & 0.15 & 0.13 & 0.13
		\\
		$t\bar t$& 5403 & 8.81 & 2.68 & 0.13 & 0.006
		\\
		$WW$ & 724.1 & 0.55 & 0.41 & 0.006 & 0.006
		\\
		$tW$ & 320.4 & 1.25 & 0.19 & 0.008 & 0.002 
		\\
		\hline
	\end{tabular}
	\caption{Signal and background cross sections at 13 TeV (in fb) after
	each cut.  $p_T(l_1)$ and $p_T(l_2)$ are the $p_T$s of the leading and
	the trailing leptons, respectively.\label{tab:cuts}}
\end{table*}

In the scalar sector, the production cross section of $\eta$
through gluon-fusion is $6$ fb at $13 \tev$ at the leading order, while
the next-to-leading order QCD correction will raise it to $\sim10$ fb.
However, it decays into gluons with
a $\sim98\%$ branching ratio (BR), thus making its discovery particularly difficult \cite{Sirunyan:2016iap}. The second
largest decay process of $\eta$ is into photons, with $\mathrm{BR}\sim0.4\%$, that could
become a valid search for HL-LHC.
Lastly, the production cross section of a pair of $\omega$'s is completely
irrelevant at LHC energies, while a $100 \tev$ collider could produce it
with a $\approx5$ fb cross section.  For our benchmark parameters, 
$\omega$'s could decay into a pair of $\Lambda$'s.  Thus, for the
di-$\omega$ production channel one of the most interesting final states will be four
hard leptons plus large MET.  This provides a distinct and
characteristic signal that can help to discriminate our model from other new
physics scenarios, and is essentially free of background.
All the cross sections, decay rates, and branching ratios,
presented in this section, were obtained using {\sc
MadGraph5\_aMC@NLO} \cite{Alwall:2014hca} and {\sc MadWidth}
\cite{Alwall:2014bza} with a UFO model
\cite{Degrande:2011ua} made with the {\sc FeynRules} package
\cite{Alloul:2013bka}.  Loop-induced processes were computed following
Ref.~\cite{Hirschi:2015iia}, and the corresponding counter term was computed with
{\sc NLOCT} \cite{Degrande:2014vpa}.

\section{Conclusions}
In this letter, we presented a simple model addressing two major problems of cosmology, 
baryogenesis and the ratio of DM to baryonic matter energy densities, and consequently predicting the mass of DM.  The model is an extension of the one presented in Ref.~\cite{exo-Higgs}, with the addition of a complex scalar $\chi$, that carries a conserved global charge $Q_\chi$. Requiring that $\chi$ is lighter than all the other particles carrying $Q_\chi$, we ensure that $\chi$ is a stable particle and a good asymmetric DM candidate.   

It is convenient here to summarize the main steps of our scenario.
Baryogenesis happens at the temperature of the exo-symmetry phase transition that, being strongly first order, provides one of the Sakharov conditions \cite{Sakharov:1967dj}. The exo-CKM for the $\Qoppa$ fermions can be easily set to be the source of the $CP$ violation, and lastly a net $B-L$ number is generated by the exo-sphalerons. During this process a net $Q_\chi$ number is also generated that, after all the other particles have decayed, is completely stored in $\chi$ particles. This gives us the ratio between $Q_\chi$ and $B$ number, 
and equivalently a prediction for the DM mass 
$m_\chi\approx 1.3 \gev$.

Our model of baryogenesis  yields a phenomenology characterized by some interesting features that could help uncover some of its key assumed interactions.  In particular, the economical choice of the DM Yukawa couplings for both the transfer of asymmetries to the SM and symmetric DM population annihilation leads to the prediction that the exo-leptons are quite possibly within the reach of the LHC in the coming years.    These states can be produced in Drell-Yan processes and yield distinct signals marked by significant missing energy accompanied by charged lepton pairs. The requisite strong first order phase transition in our model can also lead to gravitational wave signals at envisioned observatories. Assuming UV extensions, our setup can result in the modification of the muon $g-2$, mediated by the DM particle.  Such a deviation may have already been observed and could be confirmed by a new measurement at Fermilab in the next few years \cite{Gray:2015qna}.

\section*{Acknowledgements}
This work is supported by the U.S. Department of
Energy under Grant Contracts DE-SC0012704. C.Z.~is partly supported by the 100-talent project of 
Chinese Academy of Sciences.

\appendix

\section{Appendix}
Here we consider a simple extension of the model presented in this letter that could be a good UV completion for the dimension 6 operator in Eq.~(\ref{op}).
Consider a new vector-like $SU(2)_L$ doublet $F$ and a new vector-like singlet $\psi$ with charge $Q_\chi =+1$ whose masses $m_F\sim m_\Psi\sim$ few TeV are generated by some high scale physics. 
The interactions of these fermions with the leptons in the exo-SM sectors are given by the Yukawa terms
\bea
&& \lambda_F \chi \bar{F}_R \ell_L+ \lambda_\psi \chi \bar{\psi}_L \ell_R\\
&+& (\bar{\Lambda}_L,\bar{F}_L,\bar{\psi}_L)
\left(
\begin{array}{ccc}
m_\Lambda& y_1 H &  m_{\psi\Lambda} \\
 0 & m_F  & y_2 H  \\
y' \eta & y_3 H  & m_\psi  
\end{array}
\right)
\left(
\begin{array}{c}
\Lambda_R \\
F_R  \\
 \psi_R
\end{array}
\right),\nonumber
\label{UV1}
\eea
where for simplicity we suppressed the family indices. $m_{\psi\Lambda} $ can in principle be generated by the same physics that gives masses to $F$ and $\psi$. 
As a benchmark, we use Eq. (\ref{eq:benchmark}) with the appropriate modifications/additions: we take $m_\Lambda= 1.4$ TeV, so that after diagonalization its mass is $\sim 1$ TeV, $m_F=3.5$ TeV, $m_\psi=4$ TeV, $m_{\psi\Lambda}=0.6$ TeV and $y_i=0.2$ and $y' = 1$.
Thus the correction $\delta a_\ell$ to the leptonic $g-2$ for 
$\lambda_\ell=\lambda_F=1$ and $\lambda_\psi=0$ is 
\beq
\delta a_\ell\sim 2.6\times 10^{-9}.
\eeq
Finally, one can ask if these new fields can destabilize the EW vacuum or push the weak coupling to become non-perturbative, since they couple to both SM Higgs and $W$ bosons.  One can easily check that the above model does not lead to a change in the sign of the $SU(2)_L$ coupling $\beta$ function.  Also, new Yukawa couplings of $\sim 0.2$ would not affect the vacuum stability of the SM Higgs appreciably.

\end{document}